\documentclass[%
 aip,
 amsmath,amssymb,
 reprint,%
]{revtex4-1}

\usepackage{graphicx}
\usepackage{dcolumn}
\usepackage{bm}

\usepackage[utf8]{inputenc}
\usepackage[T1]{fontenc}
\usepackage{mathptmx}
\usepackage{etoolbox}

\makeatletter
\def\@email#1#2{%
 \endgroup
 \patchcmd{\titleblock@produce}
  {\frontmatter@RRAPformat}
  {\frontmatter@RRAPformat{\produce@RRAP{*#1\href{mailto:#2}{#2}}}\frontmatter@RRAPformat}
  {}{}
}%
\makeatother
\begin{document}

\preprint{AIP/123-QED}

\title[Simulation studies of Single Event Effects in $\beta$-Ga$_2$O$_3$ MOSFETs]{Simulation studies of Single Event Effects in $\beta$-Ga$_2$O$_3$ MOSFETs}
\author{Animesh Datta}
\author{Uttam Singisetti}%
 \email{uttamsin@buffalo.edu}
  \email{animeshd@buffalo.edu}
\affiliation{ 
Department of Electrical Engineering, University at Buffalo, The State University of New York, Buffalo, New York 14228, USA
}%

\date{\today}

\begin{abstract}
In this article, we investigate the Single Events Effects (SEE) leading to Single Event Burnout (SEB) in $\beta$-Ga$_2$O$_3$ MOSFETs. Using Silvaco TCAD, 2D simulations were performed to understand the mechanism behind the SEB mechanism in lateral Ga$_2$O$_3$ MOSFETS. The high electric fields in the channel played a critical role leading to high impact generation rates and eventual SEB. To reduce the electric field in the channel, radiation hardened designs are then proposed with rounded gates and the use of a combination of high permittivity (k) dielectric with SiO$_2$. With HfO$_2$-SiO$_2$ dielectric combination, the SEB threshold of 550V at LET=10 MeV/mg/cm$^2$ is seen. However, to operate under extreme radiation conditions, a combination of very high-k dielectric material BaTiO$_3$ with SiO$_2$ is proposed. Using the radiation hardened design, SEB thresholds up to 1000 V for LET=75 MeV/mg/cm$^2$ could be achieved which is higher than the state-of-the-art technology. The energy dissipated during the ion strike event is also calculated and it is observed that it is lower than that of SiC MOSFETs.
\end{abstract}

\maketitle

\section{Introduction}
\label{sec:introduction}
Radiation damages are a common reliability issue for power electronic devices used for space applications. Harsh radiation conditions in space can cause reliability problems such as temporary loss of data, circuit degradation, loss of functional operation to destruction of the semiconductor device\cite{srour2006framework,dodds2015contribution,martinez2018radiation}. Radiation effects are typically classified into cumulative effects such as Total Ionizing Dose (TID) effects, or Single Event Effects (SEE) leading to Single Event Burnout (SEB). These effects lead to defect generation, device degradation, and device failures. Previous studies have suggested that SEE effects are more dominant than cumulative effects in wide band-gap materials\cite{kim2019radiation,look1997defect,ionascut2002radiation,barry1991energy}. There have been significant experimental and simulation studies of radiation effects especially single event effects in SiC\cite{witulski2017single,li2021high,mcpherson2018heavy,javanainen2016heavy,mizuta2014investigation,makino2013heavy,kuboyama2006anomalous,ball2019ion} and GaN\cite{pearton2015radiation,lee2017defects,yadav2015low,patrick2015simulation} devices. Recent reports have also explored SEE in ultrawide bandgap AlGaN HEMTs \cite{liu2022simulation}and MISFETs \cite{luo2019research} to design radiation hardened devices.

$\beta$-Ga$_2$O$_3$ with its large band gap of 4.8-4.9 eV and high critical electric field strength of 8 MV/cm\cite{tsao2018ultrawide,tadjer2016structural} has promising potential for power electronics and RF applications. Another advantage of $\beta$-Ga$_2$O$_3$ is the availability of large diameter wafers. Common growth techniques like Czochralski (CZ), float-zone (FZ), edge-defined film fed (EFG), or Bridgman can be used to grow bulk crystals\cite{baldini2016si,galazka2016scaling,stepanov2016gallium}. Recent experimental reports have shown high breakdown voltages (multi-kVs) and high Figure of Merit in $\beta$-Ga$_2$O$_3$ lateral MOSFETs\cite{9499054,bhattacharyya20224,sharma2020field,zeng2019field,mun20192,chabak2016enhancement,lv2020lateral} and Schottky diodes\cite{li20181,schubert2020gallium,zhang2022ultra,9926082,9456924}. Previously there have been studies of neutron damage, x-ray damage, and gamma-ray damage in $\beta$-Ga$_2$O$_3$ MOSFETs\cite{kim2019radiation,wong2018radiation,lingaparthi2019surface,polyakov2022deep,yakimov2021experimental} but there has been no extensive study on single event effects in these devices despite it being a favorable ultra-wide band gap semiconductor. 

In this work, we investigate the SEE effects in  $\beta$-Ga$_2$O$_3$ MOSFETs using 2D TCAD simulations. Initially, a baseline design of lateral $\beta$-Ga$_2$O$_3$ MOSFET with SiO$_2$ dielectric is simulated under various radiation conditions. The SEB threshold of 200V for LET=10 MeV/mg/cm$^2$ for this device is lower than the simulated SEB of AlGaN/GaN channel MISFETs which has a threshold of 430V at  LET=21 MeV/mg/cm$^2$  \cite{luo2019research}. The physics behind the SEB mechanism is investigated which helps us in proposing radiation hardened designs. The simulations suggest that the use of a combination of high-k dielectric with SiO$_2$ is crucial for full device recovery under extreme radiation conditions. Using a conventional HfO$_2$ dielectric increases the SEB threshold voltage from 220V to 550V at LET=10 MeV/mg/cm$^2$ which is higher than that of GaN channel HEMTs\cite{liu2022simulation}. But this design is still susceptible to SEB at higher LETs and operating biases. Recent reports have suggested the use of a BaTiO$_3$ on $\beta$-Ga$_2$O$_3$ to achieve high breakdown voltage devices\cite{xia2019metal}. There have also been reports of using BaTiO$_3$ on AlGaN to design state of the art breakdown voltage diodes\cite{razzak2020batio3} and HEMTs\cite{rahman2021integration}. BaTiO$_3$ has a reported dielectric constant of $>$100 for thin films. Thus a design with an extreme high-k dielectric (BaTiO$_3$) in combination with SiO$_2$ is proposed which shows full recovery for all radiation conditions. With this design, SEB thresholds of 1000 V at LET=75 MeV/mg/cm$^2$ can be achieved which is higher than the state of the art technology.
\section{Device Structure and Models}
\label{sec:DSM}
\begin{figure}[h]
\begin{center}
    \includegraphics[width=\linewidth]{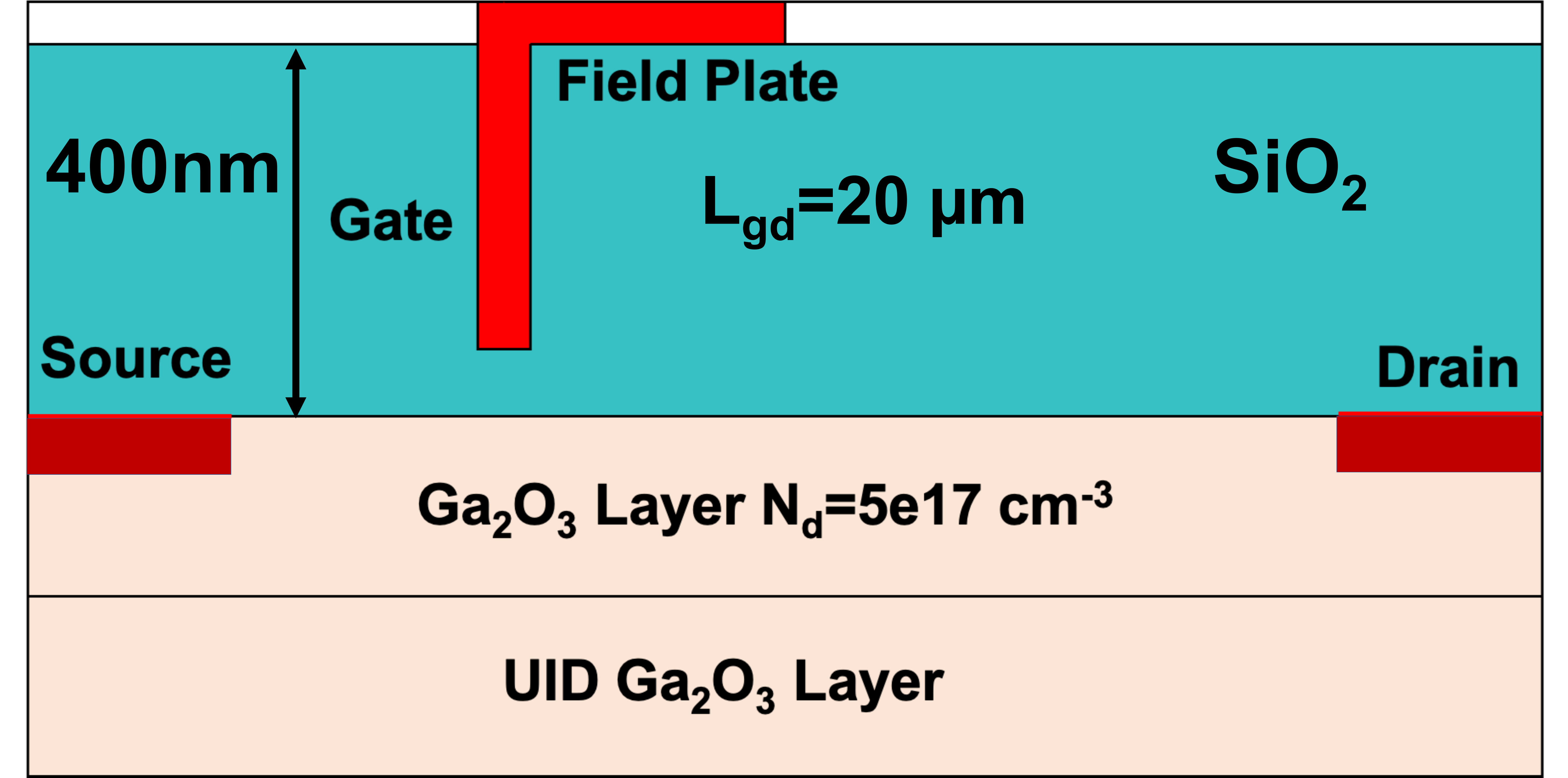}
    \caption{Device schematic of the simulated baseline lateral MOSFET}
    \label{fig1}
    \end{center}
\end{figure}
\begin{figure}[h]
\begin{center}
    \includegraphics[width=\linewidth]{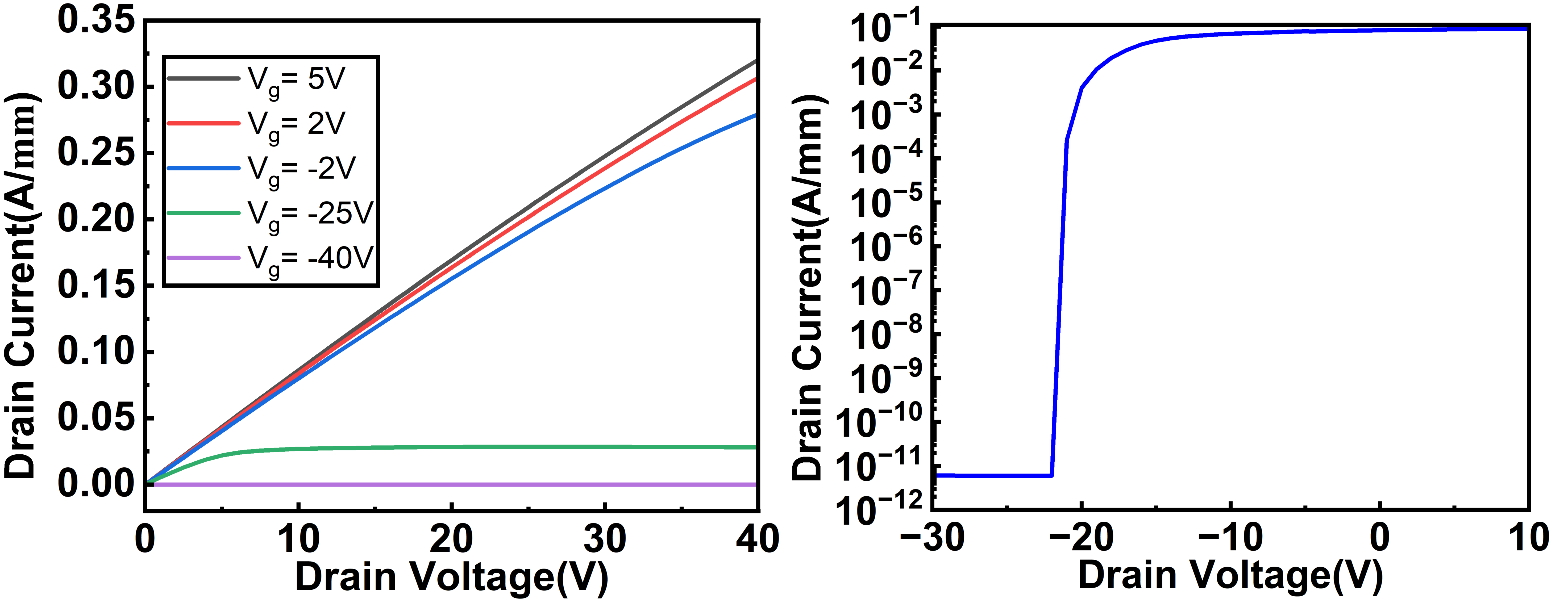}
    \caption{Simulated (a) output characteristics and (b) Transfer characteristics of the MOSFET}
    \label{fig2}
    \end{center}
\end{figure}
\begin{figure}[h]
\begin{center}
    \includegraphics[width=\linewidth]{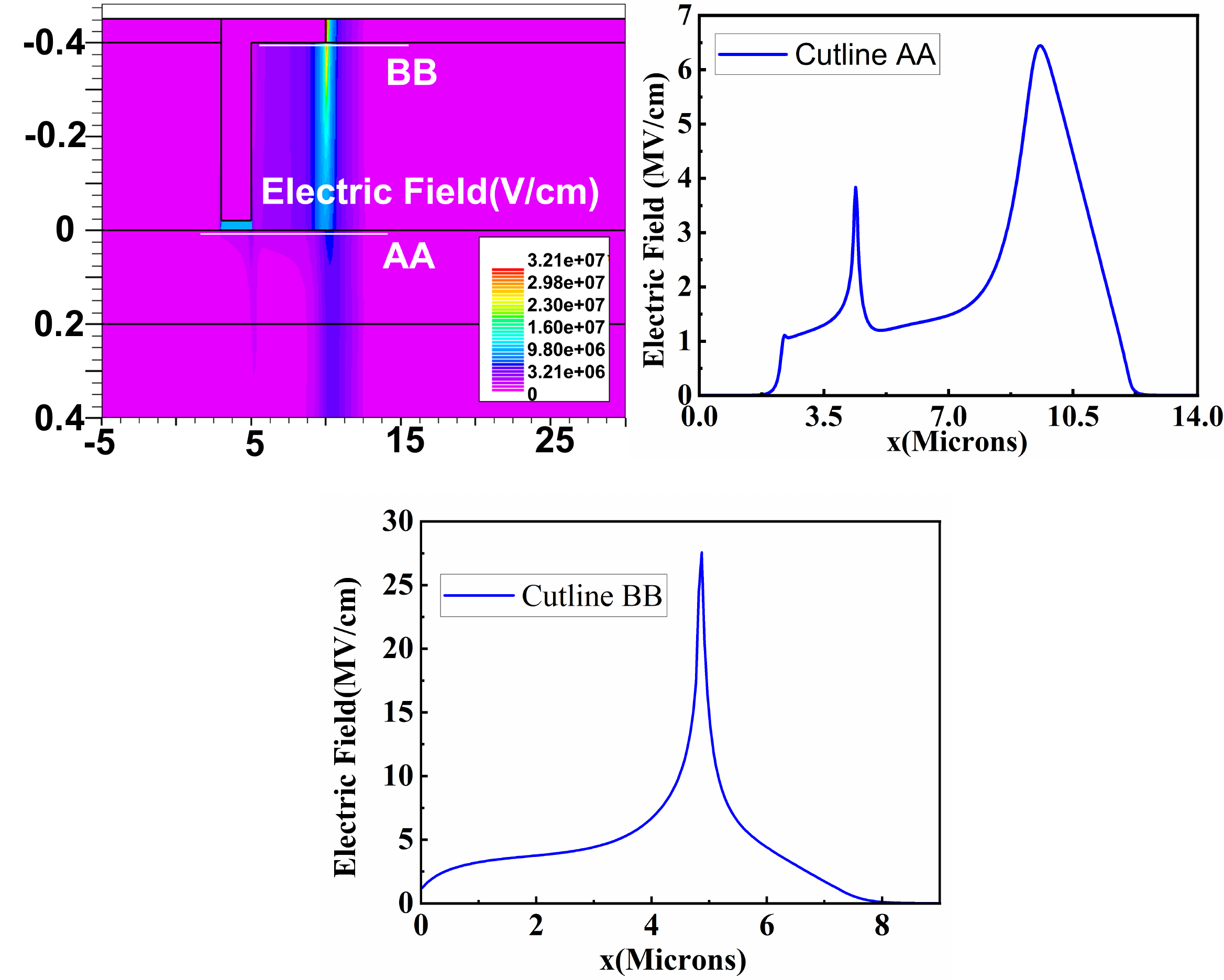}
    \caption{(a) Silvaco ATLAS simulation of electric field profile at 1500V (b)Field in the Ga$_2$O$_3$ channel along cutline AA' and (b) Field in the field plate oxide along cutline BB'}
    \label{fig3}
\end{center}
\end{figure}

Fig~\ref{fig1} shows the device structure of the baseline Ga$_2$O$_3$ MOSFET. The channel layer is 200 nm thick with a doping concentration of 5$\times$10$^{17}$ cm$^{-3}$. The ohmic layers, used for the drain and source contacts are 50 nm thick and heavily doped with a concentration of 1$\times$10$^{19}$ cm$^{-3}$. The gate electrode has a work function of 4.8 eV. Field plate edge termination technique \cite{sharma2020field,zeng2019field} is used to maximize the breakdown voltage of the device. The device was simulated using the 2-D simulator of SILVACO ATLAS. The models used in the simulation include the SRH recombination model, Auger recombination model, impact ionization model, and the field dependent mobility model. According to previous studies from first principle calculations\cite{ghosh2016ab,ghosh2017electron} the low field mobility of Ga$_2$O$_3$ is taken to be 150 cm$^2$/V-s for a doping concentration of 5$\times$10$^{17}$ cm$^{-3}$. The hole mobility in the channel is taken to be 0.1 cm$^2$/V-s which is quite low due to the weakly interacting O 2p states which create deep acceptors. The impact ionization parameters for $\beta$-Ga$_2$O$_3$ have been taken from a detailed first principle theoretical study\cite{ghosh2018impact}. 
Fig.~\ref{fig2} shows the output and transfer characteristics of the device. The threshold voltage of the device from Fig.~\ref{fig2} is -20 V.
\\
The breakdown voltage of the device is determined by analyzing the electric field contour plots at specified operating bias. Fig.~\ref{fig3} shows the electric field profile at the breakdown voltage of the device. 
The device breakdown is determined when the electric field in the channel exceeds the critical breakdown field of SiO$_2$ and  Ga$_2$O$_3$. As shown in Fig.~\ref{fig3} at 1500 V, the field in the channel is close to 7 MV/cm and that of SiO$_2$ is 25 MV/cm which is close to their critical breakdown fields. Thus the breakdown voltage of the device is estimated to be 1500V, which is close to the experimentally reported breakdown voltage \cite{sharma2020field,zeng2019field} at comparable gate-drain spacing. 
\section{Single Event Effects}
\label{sec:SEE}
Single Event Effects (SEE) occur when individual energetic particle induces errors or failures in the device or the overall circuit\cite{petersen2011single,witulski2018single,khachatrian2016application}. The incident ionization particle loses energy in the semiconductor through Coulomb interaction with the lattice structure. The energy is transferred to the lattice as an ionization tail of free electron-hole pairs leading to transient current. Under heavy ion radiation conditions, a second breakdown mechanism might occur leading to a catastrophic failure of the device known as Single Event Burnout (SEB).

In radiation studies, a common measure of the loss of energy of the SEU particle, as it suffers collisions in a material, is the linear energy transfer (LET) value, which is given in units of MeV/mg/cm$^2$. LET is the linear energy transfer of the incident ion which determines the number of electron hole pairs (ehp) created in the material depending on the band structure and the density of the material. The creation energy of an electron-hole pair in the semiconductor($\epsilon_1$) is approximately calculated to be three times the band gap of the material\cite{alig1975electron}. The number of free carriers released is calculated as\cite{weatherford2003radiation}: 
\begin{equation}
    LET(ehp/\mu m)=LET(MeV\frac{mg}{cm^2})\times\frac{1}{\epsilon_1}\times density(gm/cm^3)
\end{equation}
 The amount of electron-hole pairs created by the ionizing particle can be converted in terms of the linear charge deposited inside the semiconductor in units of pC/$\mu$m. Since Ga$_2$O$_3$ is a wide band gap semiconductor, the conversion factor for material comes to be 0.007 pC/$\mu$m corresponding to LET= 1 MeV/mg/cm$^2$. This is quite low compared to 0.1 pC/$\mu$m for Si and is comparable to other wide band gap materials like GaN and SiC. Thus the lower charge deposition by the ionizing particle and the highly promising properties Ga$_2$O$_3$ based devices might provide better radiation tolerance than other wide band gap materials.
\\
\\
The single event effects can be simulated using the ATLAS 2D simulator by using the SINGLEEVENTUPSET statement. The radius, length, and time dependence of the charge generation track can be specified. The coordinates of the entry and exit points of the track can also be specified within the device. In 2D simulations, the track is assumed to be a cylinder with a specified radius. The electron/hole pairs generated at any point are a function of the radial distance \textit{r}, from the center of the track to the point, the distance \textit{l} along the track, and the time, \textit{t}. ATLAS implements the generation rate as the number of electron-hole pairs per $cm^3$ along the track according to the equation:
\begin{multline*}
    G(r,l,t)=(DENSITY*L1(l) + \\
    S*B.DENSITY*L2(l))*R(r)*T(t)
\end{multline*}
where DENSITY and B.DENSITY are defined as the number of generated electron/hole pairs per cm$^{-3}$. By setting a PCUNITS parameter, we can specify the charge generated in terms of pC/$\mu m$ depending on the LET value of the ionizing radiation and the concerned semiconductor as discussed above. The factors L1 and L2 define the variation of charger or carrier generation along the SEU track. The factor R(r) is the radial parameter and is defined as:
\begin{equation}
    R(r)=exp(-\frac{r}{RADIUS})^2
\end{equation}
The factor \textit{T(t)} is the time dependency of the charge generation governed by two user-defined parameters namely the temporal Gaussian function width and the initial time of charge generation.

\section{SEE Simulations}
The single event effects in Ga$_2$O$_3$ MOSFETs are investigated using the single event upset model in SILVACO TCAD. In accordance with the previous radiation research studies\cite{ball2019ion,luo2019research}, the spatial Gaussian function width is set to 50 nm, the temporal Gaussian function width is set to 2 ps, and the initial time of charge generation is set to 100 ps. The strike location is chosen by studying the electric field profile of the device before radiation. As shown in Fig.~\ref{fig2}b, the location in the channel vertically under the edge of the field plate has a very high peak electric field and thus is the most sensitive region for SEE events following an ion strike. Thermal simulation models are also included but our analysis suggests that the temperature of the device stays within the tolerance limits. The simulation parameters along with the device parameters are shown in Table.~\ref{table1}.

\begin{table}[]
\centering
\caption{Design Parameters in simulation}
\label{table1}
\resizebox{\columnwidth}{!}{
\begin{tabular}{|l|l|}
\hline
\textbf{Design Parameters}           & \textbf{Value}             \\ \hline
Lgd(Gate to Drain Length)            & 20 $\mu m$                 \\ \hline
Thickness of dielectric              & 400 nm                     \\ \hline
Channel Layer doping                 & 5$\times$17 cm$^{-3}$      \\ \hline
Dielectric constant of SiO2          & 3.9                        \\ \hline
Dielectric constant of HfO2          & 25                         \\ \hline
Dielectric constant of BaTiO3        & 100                        \\ \hline
Spatial Gaussian function width      & 50 nm    \cite{ball2019ion,luo2019research}                  \\ \hline
Temporal Gaussian function width     & 2$\times$10$^{-12}$s \cite{ball2019ion,luo2019research}                    \\ \hline
Initial Time of of charge generation & 100e-12  \cite{ball2019ion,luo2019research}                  \\ \hline
Length of ion track                  & 10,0 - 10,0.4              \\ \hline
Hole recombination time              & 4$\times$10$^{-10}$s        \\ \hline
Models used                          & auger, srh, fldmob, impact,lat.temp \\ \hline
Impact Ionization Parameters.        & a=0.79$\times$10$^6$, b=2.92$\times$10$^7$ \cite{ghosh2018impact} \\ \hline
\end{tabular}
}
\end{table}

\begin{figure}[h]
\begin{center}
    \includegraphics[width=\linewidth]{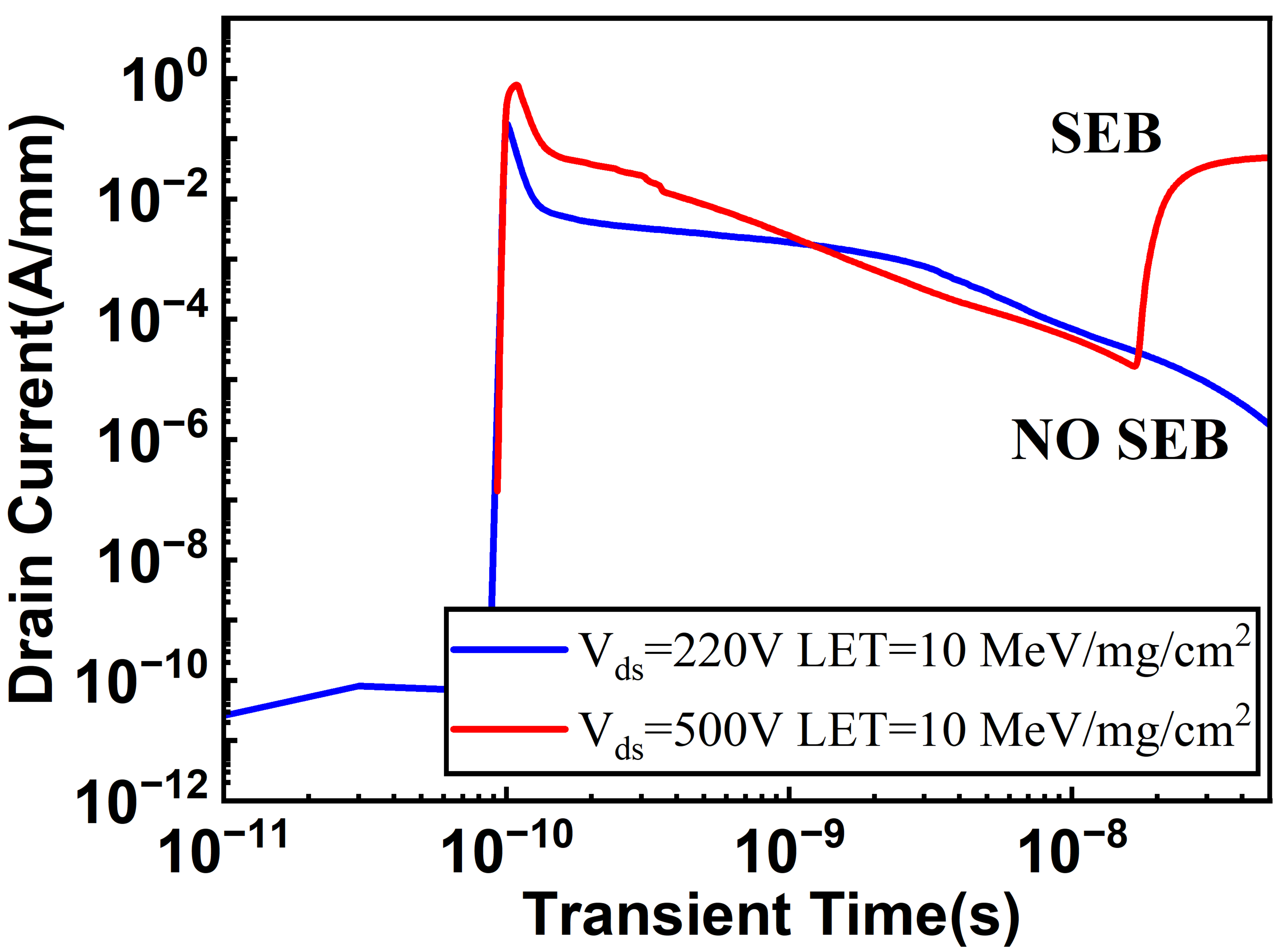}
    \caption{Transient drain current for the MOSFET at V$_{ds}$=220 V and 500 V for LET=10 MeV/mg/cm$^2$}
    \label{fig4}
    \end{center}
\end{figure}

\begin{figure}[h]
\begin{center}
    \includegraphics[width=\linewidth]{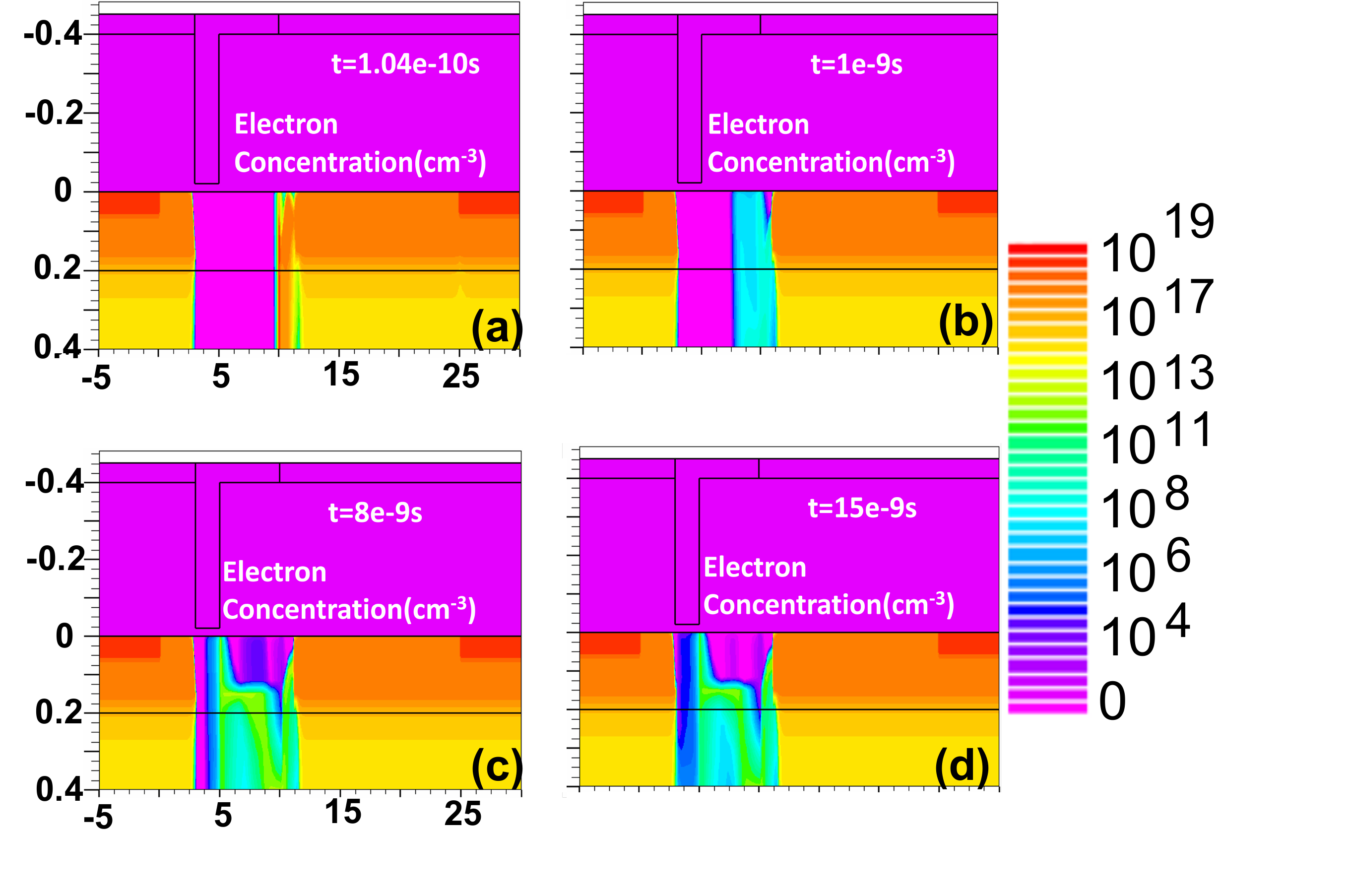}
    \caption{Electron concentration at various time instants after the ion strike at V$_{ds}$=500 V and LET=10 MeV/mg/cm$^2$}
    \label{fig5}
    \end{center}
\end{figure}

\begin{figure}[h]
\begin{center}
    \includegraphics[width=\linewidth]{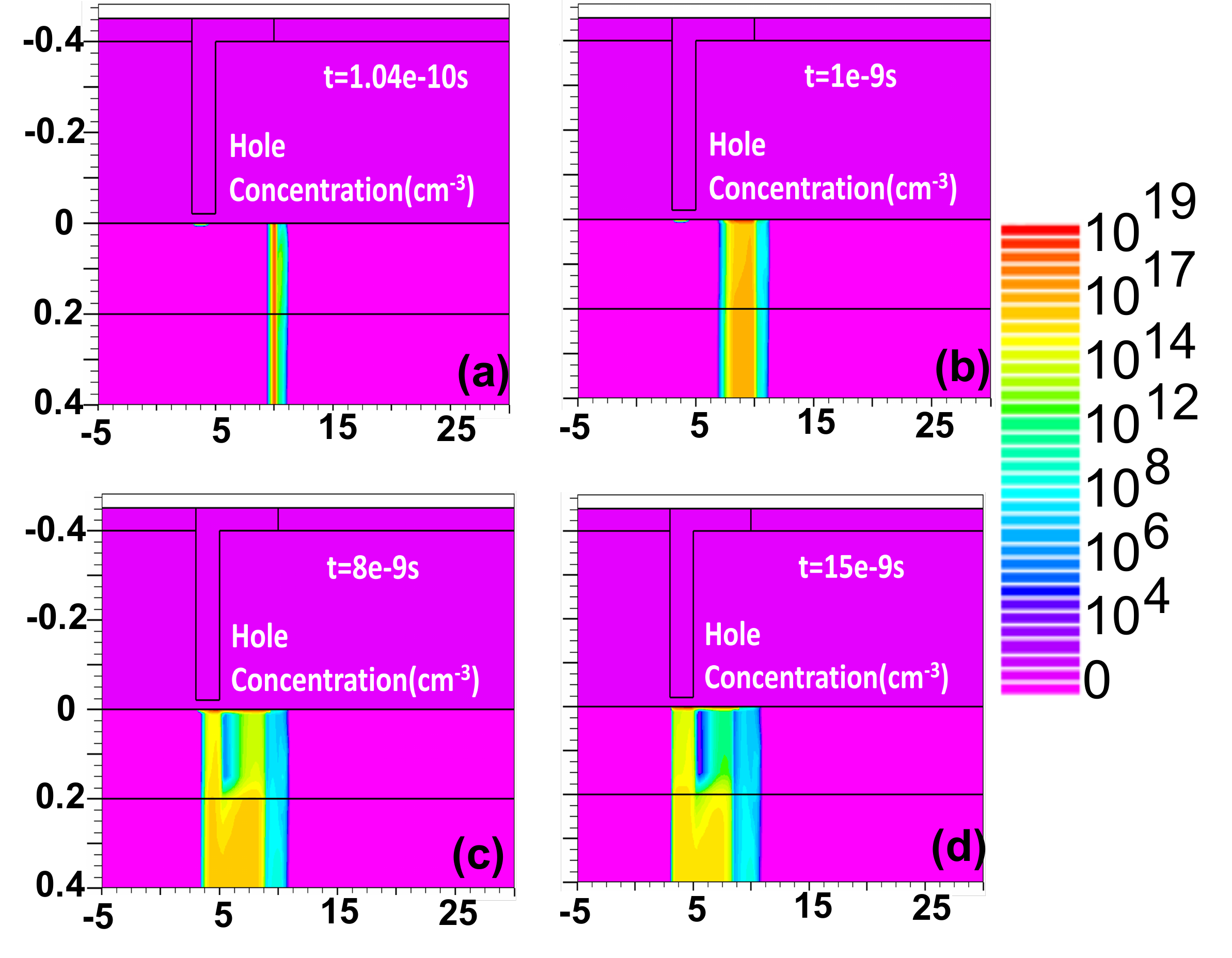}
    \caption{Hole concentration at various time instants after the ion strike at V$_{ds}$=500 V and LET=10 MeV/mg/cm$^2$}
    \label{fig6}
    \end{center}
\end{figure}

\begin{figure}[h]
\begin{center}
    \includegraphics[width=\linewidth]{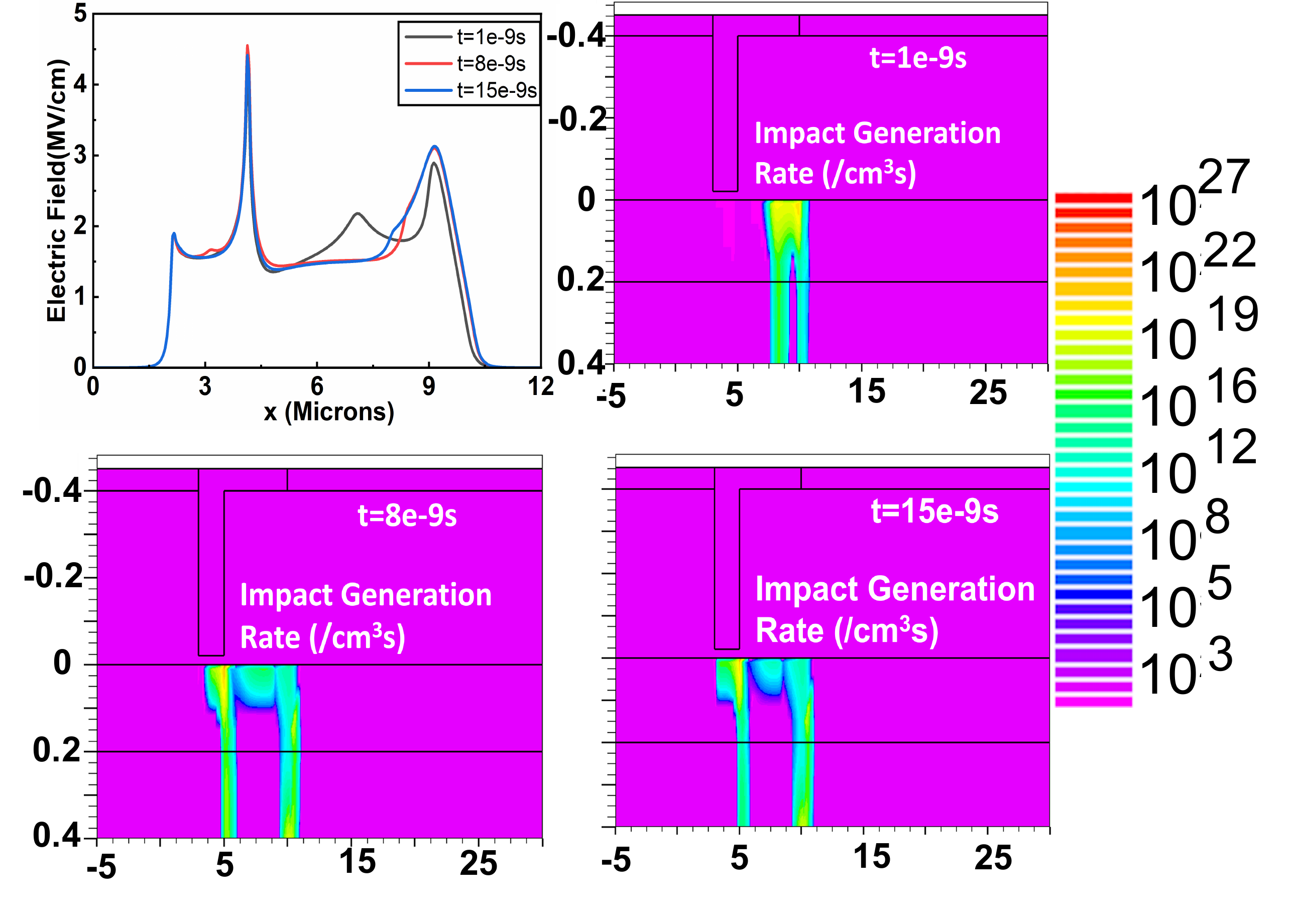}
    \caption{Electric field profile and Impact Generation Rate at various time instants after the ion strike at V$_{ds}$=500 V and LET=10 MeV/mg/cm$^2$}
    \label{fig7}
    \end{center}
\end{figure}

The MOSFET structure shown in Fig.~\ref{fig1} was simulated under two operating biases and LET=10 MeV/mg/cm$^2$ to understand the SEE effects. From Fig.~\ref{fig4} it can be seen that there is a peak transient current for both conditions just after the ion strike, but with time it slowly recovers. However, at V$_{ds}$= 500 V, a second breakdown mechanism occurs which is discussed below, the current starts to increase and goes beyond the safety limits. Thus the device recovers under V$_{ds}$=220 V but suffers from SEB at V$_{ds}$=500 V at LET=10 MeV/mg/cm$^2$. 
The physics behind the SEB failure can be understood by analyzing the electron, hole concentration profiles, and electric field contour at various time instants. Fig.~\ref{fig5} and Fig. ~\ref{fig6} show the electron and hole concentration respectively in the device at various time instants after the ion strike. At t=1.04$\times$10$^{-10}$s, which is just after the ion strike, electron-hole pairs are generated along the ion strike path as shown in Fig.~\ref{fig5}a and Fig.~\ref{fig6}a. Gradually with time, the electrons start to migrate towards the drain and the holes start to migrate towards the gate which is shown in Fig.~\ref{fig5}b and Fig.~\ref{fig6}b. Now as shown in Fig.~\ref{fig7}a, the electric field in the channel is high with peaks under the gate and at the ion strike path. With time as more and more carriers start to accumulate near the gate, the impact generation rate starts to increase gradually. As shown in Fig.~\ref{fig7}, initially the impact generation rate was present only under the ion strike path. But with time as more and more holes and electrons started accumulating under the gate, the impact generation rate increased which leads to high electron and hole concentration in the channel at t=8$\times$$10^{-9}$ onwards. At t=15$\times$$10^{-9}$, the impact generation rate is very high generating a huge concentration of electrons and holes as shown in Fig.~\ref{fig6}d and Fig.~\ref{fig7}d which are present throughout the channel. This leads to the SEB failure with current exceeding the safety limits of 1 mA/mm,\cite{liu2022simulation} as shown in Fig.~\ref{fig4}.
\\
Thus with the baseline design the SEB threshold of 220 V at LET=10 MeV/mg/cm$^2$ for the Ga$_2$O$_3$ MOSFET is lower than simulated SEB in radiation hardened GaN HEMTs\cite{liu2022simulation} and AlGaN/GaN-Based MISFET\cite{luo2019research}. 
\\
The mechanism discussed above clearly indicates that the electric field in the channel should be reduced to design radiation hardened MOSFETs. Thus a modified design with a high-k dielectric combination is proposed here.
\section{Radiation hardened design}
\begin{figure}[h]
\begin{center}
    \includegraphics[width=\linewidth]{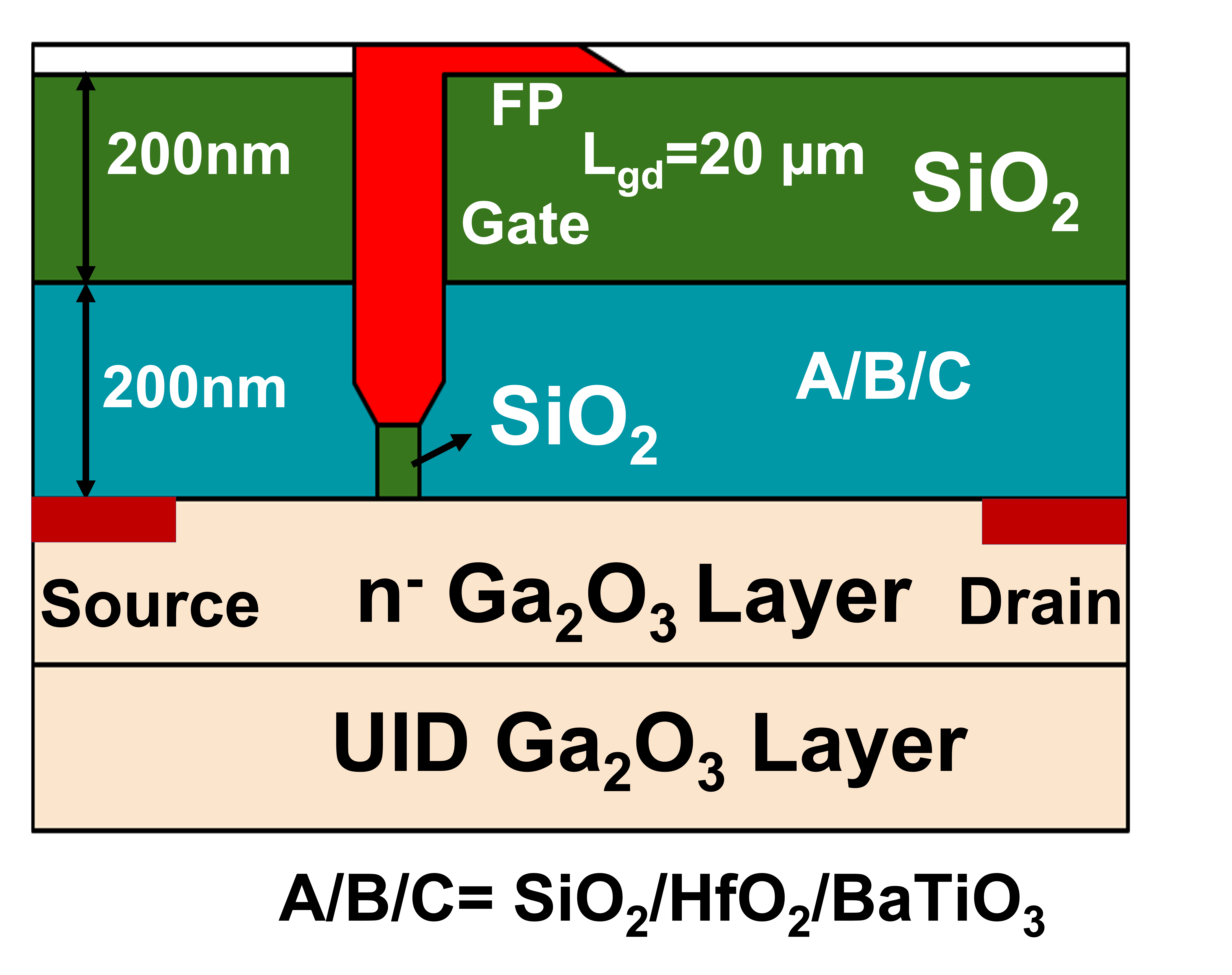}
    \caption{Schematic of radiation hardened MOSFET design}
    \label{fig8}
    \end{center}
\end{figure}

As discussed in the earlier section, to reduce the overall electric field in the channel, a MOSFET design with a combination of high-k dielectric and SiO$_2$ as shown in Fig.~\ref{fig8} is proposed.
The first design is proposed with the combination of conventional dielectrics which is HfO$_2$ and SiO$_2$. However, due to possible band alignment issues with Ga$_2$O$_3$ and HfO$_2$, the gate dielectric has been kept SiO$_2$. Gate rounding techniques have also been implemented to reduce the electric field crowding at the edges and also mimic the simulation as closely as possible to a fabricated structure. 
\\
The structure was simulated under two radiation condition of V$_{ds}$=550 V and LET=10 MeV/mg/cm$^2$ and V$_{ds}$=650 V and LET= 75 MeV/mg/cm$^2$
\begin{figure}[h]
\begin{center}
    \includegraphics[width=\linewidth]{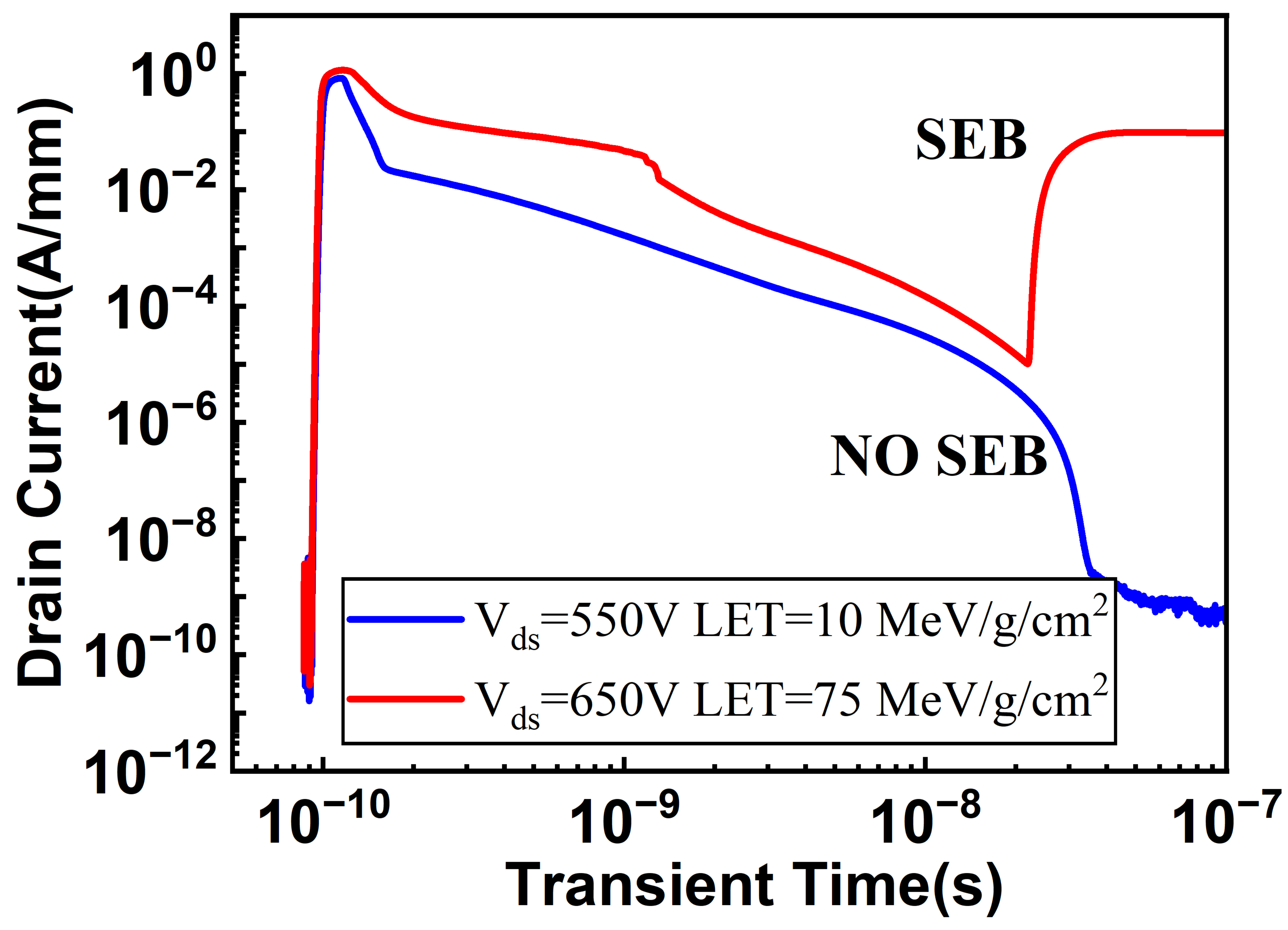}
    \caption{Transient drain current for HfO$_2$-SiO$_2$ dielectric combination MOSFET for two different radiation conditions}
    \label{fig9}
    \end{center}
\end{figure}
At V$_{ds}$=550 V and LET=10 MeV/mg/cm$^2$, the current recovers, and the leakage current is still within the safe limits indicating that the device has not suffered SEB. To understand how the proposed design helps in increasing the SEB threshold at LET=10 MeV/mg/cm$^2$, it is important to look at the electron, hole concentrations, and the impact generation rate at various time instants. During the initial period of ion strike, the electrons and holes behave similarly to as discussed in the earlier section. There is a peak transient current just after the ion strike as shown in Fig.~\ref{fig9} and then for both cases, the current starts to recover slowly. However as shown in Fig.~\ref{fig11}a, the electric field in the channel is lower especially under the gate as compared to the conventional structure though the device is operating at a higher bias. Thus at t=10$\times$10$^{-9}$s, the impact generation rate is significantly lower leading to lower levels of electron and hole concentration as shown in Fig.~\ref{fig10}. As a result, the drain current does increase again. The current stays below the safety limits and SEB is not triggered. 
\begin{figure}[h]
\begin{center}
    \includegraphics[width=\linewidth]{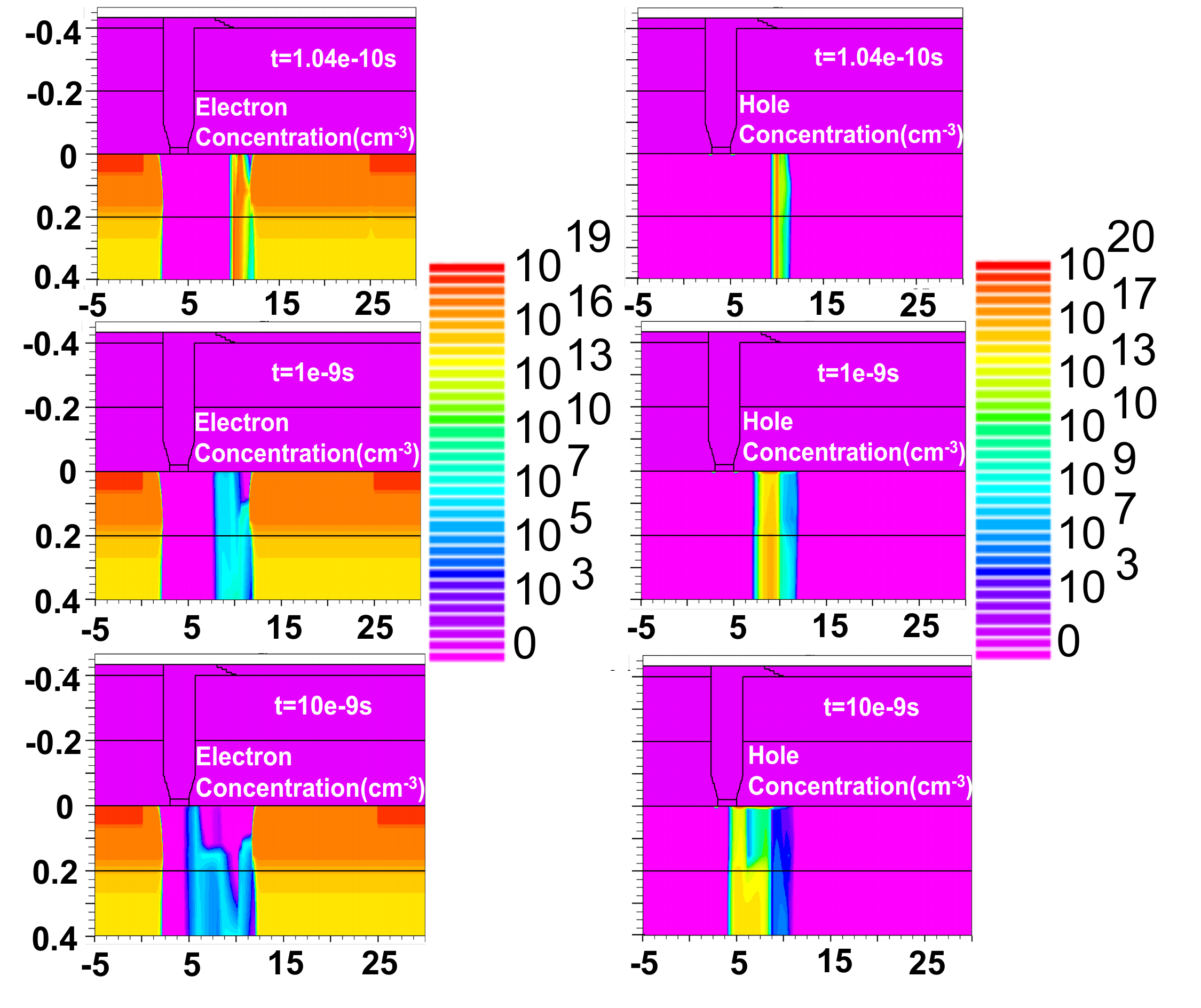}
    \caption{Electron and hole concentration at different time instants for HfO$_2$-SiO$_2$ dielectric combination at V$_{ds}$=550 V and LET=10 MeV/mg/cm$^2$}
    \label{fig10}
    \end{center}
\end{figure}

\begin{figure}[h]
\begin{center}
    \includegraphics[width=\linewidth]{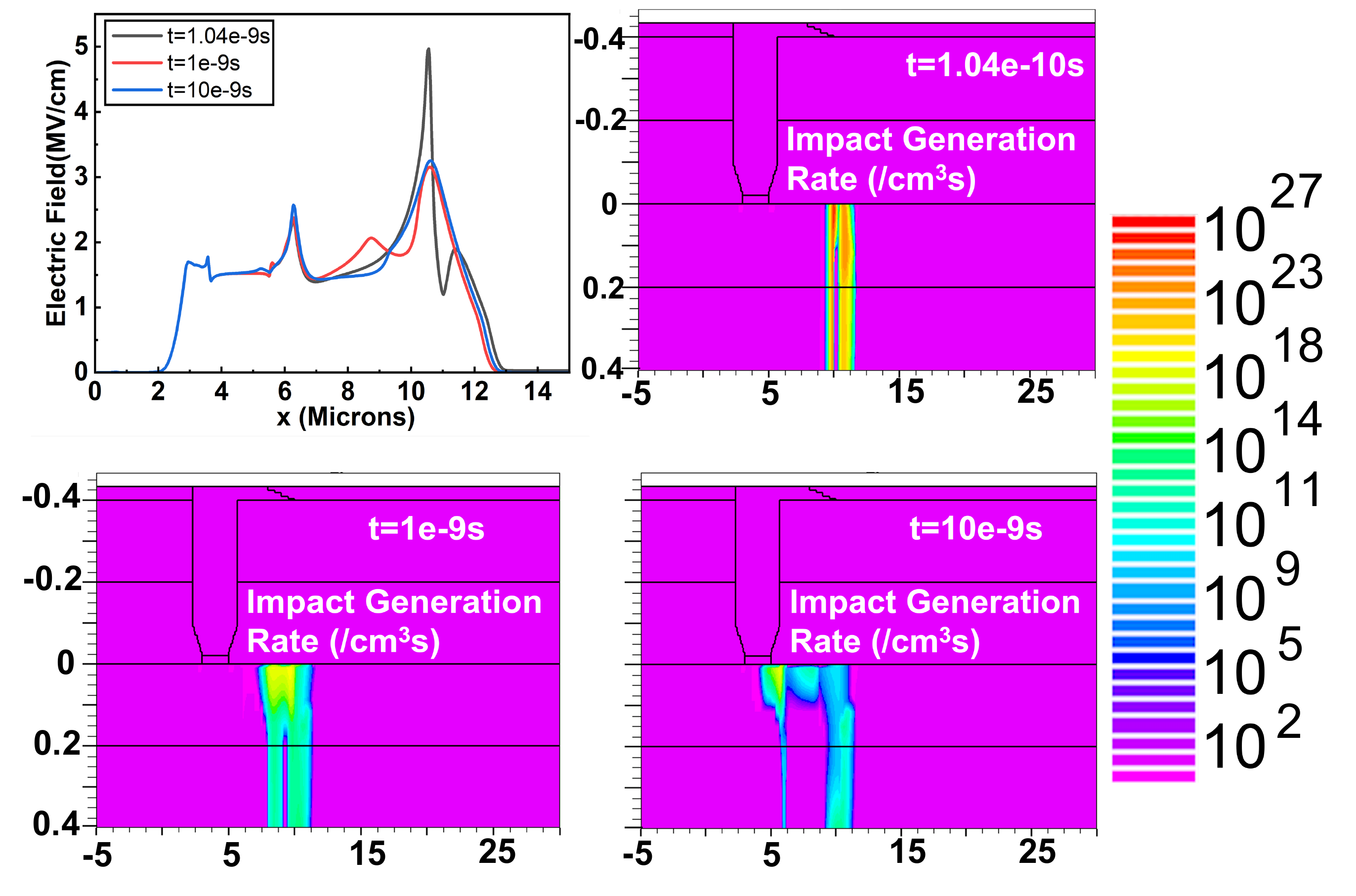}
    \caption{Electric Field and Impact Generation rate at different time instants for HfO$_2$-SiO$_2$ dielectric combination at V$_{ds}$=550 V and LET= 10 MeV/mg/cm$^2$}
    \label{fig11}
    \end{center}
\end{figure}
However, for extreme radiation conditions of V$_{ds}$=650 V and LET=75 MeV/mg/cm$^2$, the device suffers from SEB because the impact generation rate becomes higher due to higher electric fields in the channel. Thus to get further radiation hardness, a very high-k dielectric material has to be used.

\begin{figure}[h]
\begin{center}
    \includegraphics[width=\linewidth]{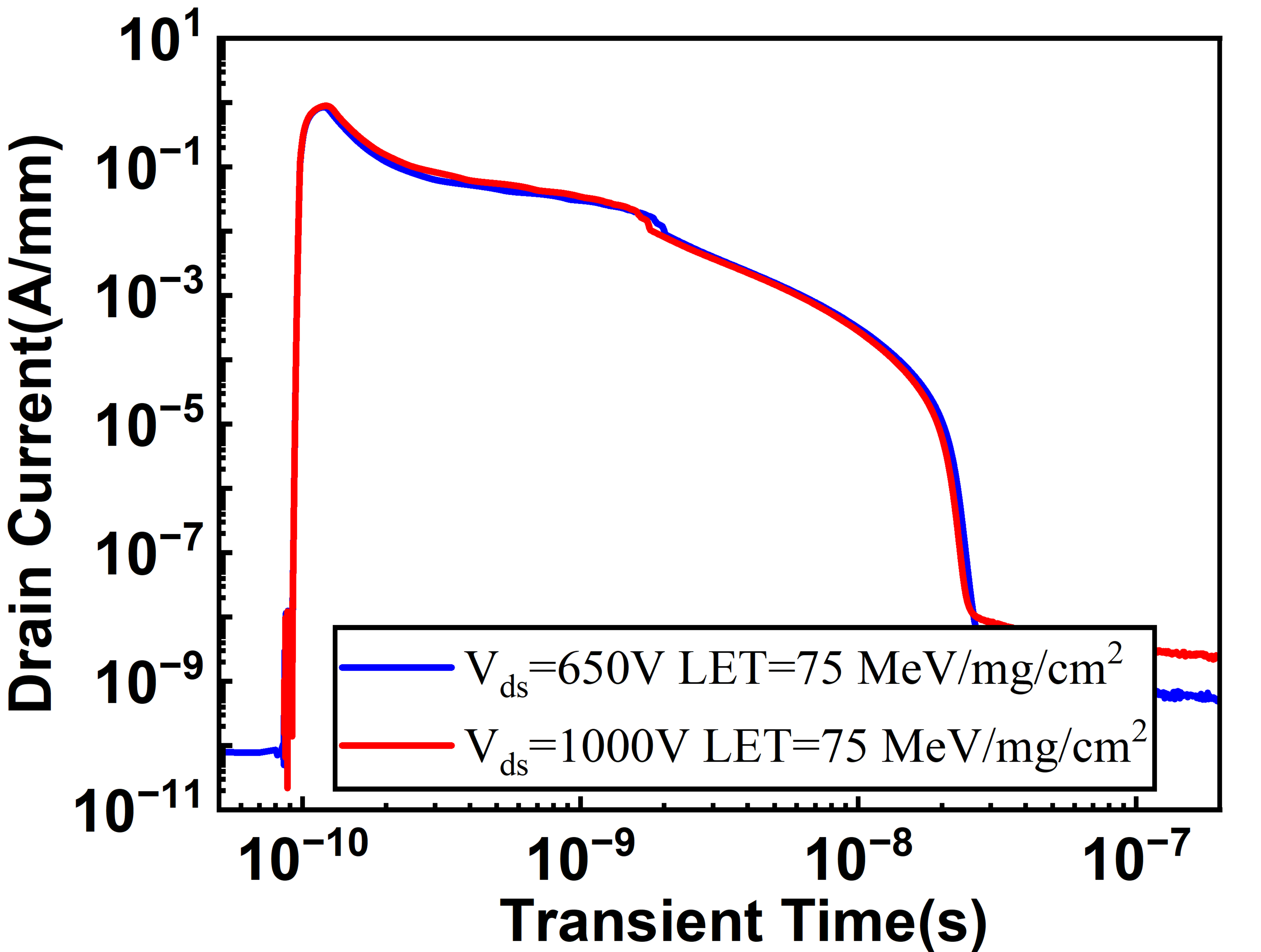}
    \caption{Transient drain current for BaTiO$_3$-SiO$_2$ dielectric combination under V$_{ds}$=650 V and LET=75 MeV/mg/cm$^2$}
    \label{fig12}
    \end{center}
\end{figure}

\begin{figure}[h]
\begin{center}
    \includegraphics[width=\linewidth]{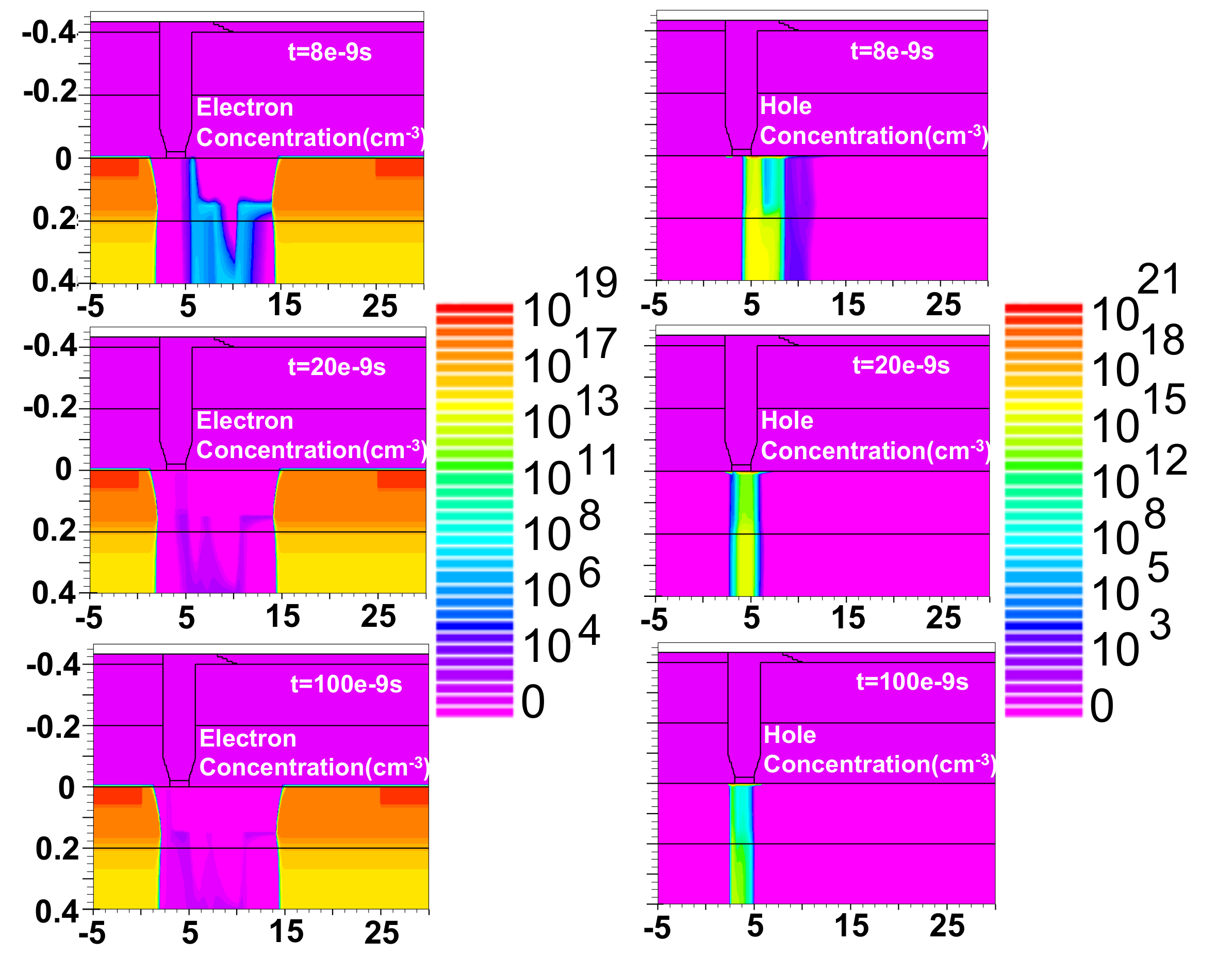}
    \caption{Electron  and hole concentration for BaTiO$_3$-SiO$_2$ dielectric combination at various time instants after the ion strike at V$_{ds}$=1000 V and LET=75 MeV/mg/cm$^2$}
    \label{fig13}
    \end{center}
\end{figure}

\begin{figure}[h]
\begin{center}
    \includegraphics[width=\linewidth]{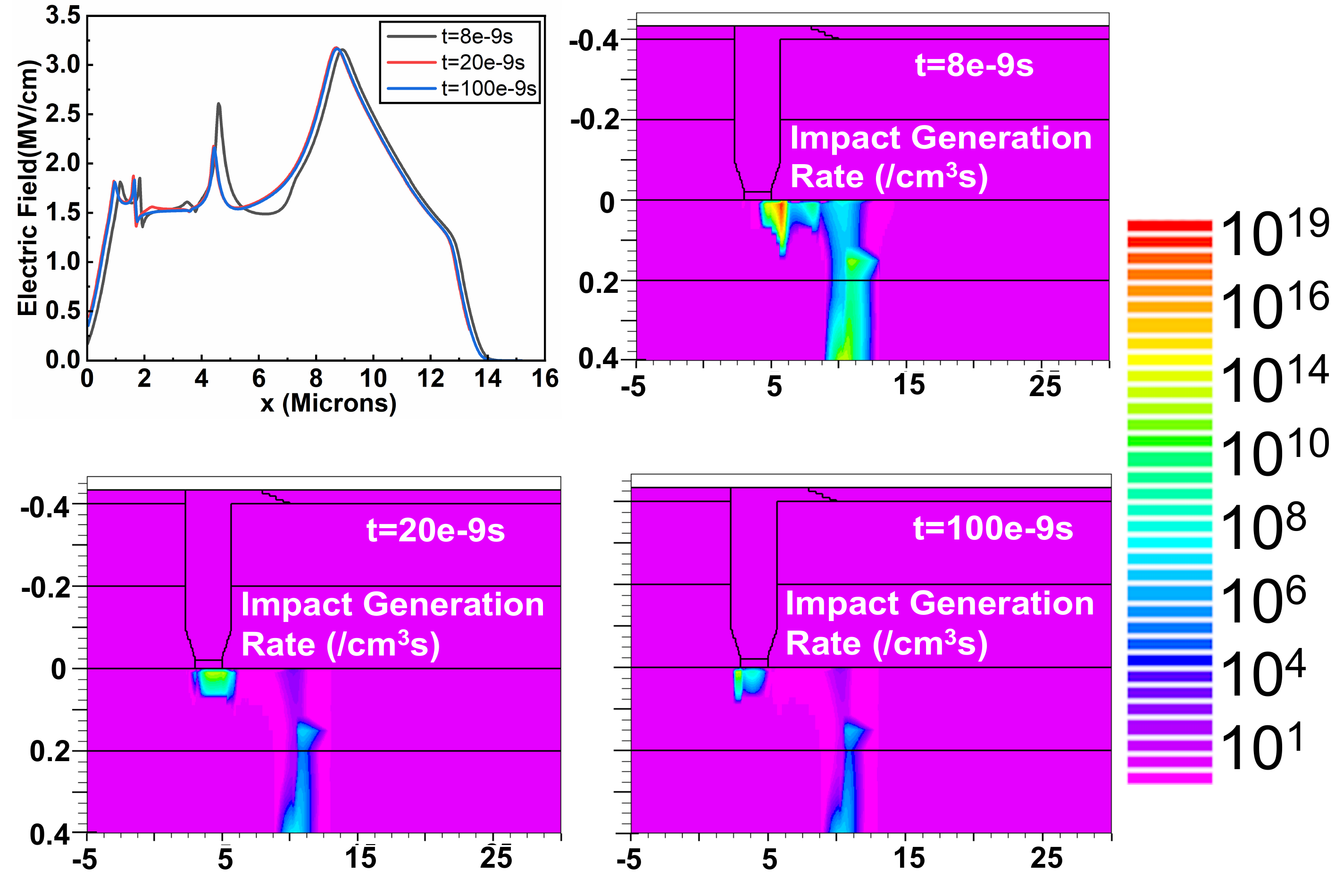}
    \caption{Electric field profile and Impact Generation Rate for BaTiO$_3$-SiO$_2$ dielectric combination at various time instants after the ion strike at V$_{ds}$=1000 V and LET=75 MeV/mg/cm$^2$}
    \label{fig14}
    \end{center}
\end{figure}

Here we have proposed a combination of BaTiO$_3$ and SiO$_2$ as shown in Fig.~\ref{fig8}. Fig.~\ref{fig12} and Fig.~\ref{fig13} shows the transient drain current plots under radiation conditions of V$_{ds}$=650 V LET=75 MeV/mg/cm$^2$ and V$_{ds}$=1000 V LET=75 MeV/mg/cm$^2$ respectively. 
The device shows full current recovery which can be further understood by analyzing the electron, hole, and impact generation rate contour plots at various time instants. As shown in Fig.~\ref{fig14}, the electric field is significantly lowered even at a high operating bias of 1000 V, which leads to low impact generation rates in the channel. The electron and hole concentration is shown from t=8$\times$10$^{-9}$s onwards since before that the behavior of electrons and holes is similar to the earlier cases. Thus even though carriers are present in the channel at 8$\times$10$^{-9}$s, they start to recombine with each other instead of undergoing impact ionization. The result is clearly seen in Fig.~\ref{fig13} where the electron and hole concentration is very low as time progresses. Therefore the drain current does not increase with time, and the device is immune to extreme radiation conditions.
\\
Ideally, if a high-k dielectric like HfO$_2$ or BaTiO$_3$ of higher thickness could be used as a fully dielectric layer for the MOSFET, the radiation hardness could be increased further. However high leakage currents and challenges of depositing thick dielectrics limit us in using that design. 

To further verify our simulations, we have also varied the hole mobility to 2 cm$^2$V$^{-1}$s$^{-1}$\cite{ma2022exploring} and have obtained similar results which shows that the hole mobility is not an important factor for determining the SEB threshold condition.
\section{Energy dissipation calculation}
The amount of energy dissipated during the ionizing radiation event in the MOSFET is an important factor in analyzing the radiation hardness of the semiconductor material. The energy dissipated during each event can be calculated by integrating the current density times area of ion track and the electric field along the entire ion track and then integrating temporally. The area of the ion track is taken to be $\pi$r$^2$, where r is the radius of the ion track. The energy dissipated is calculated using the following equation\cite{ball2019ion}
\begin{equation}
    Energy \approx \int\int J* A_{ion}*E_{field}dxdt
\end{equation}
\begin{figure}[h]
\begin{center}
    \includegraphics[width=\linewidth]{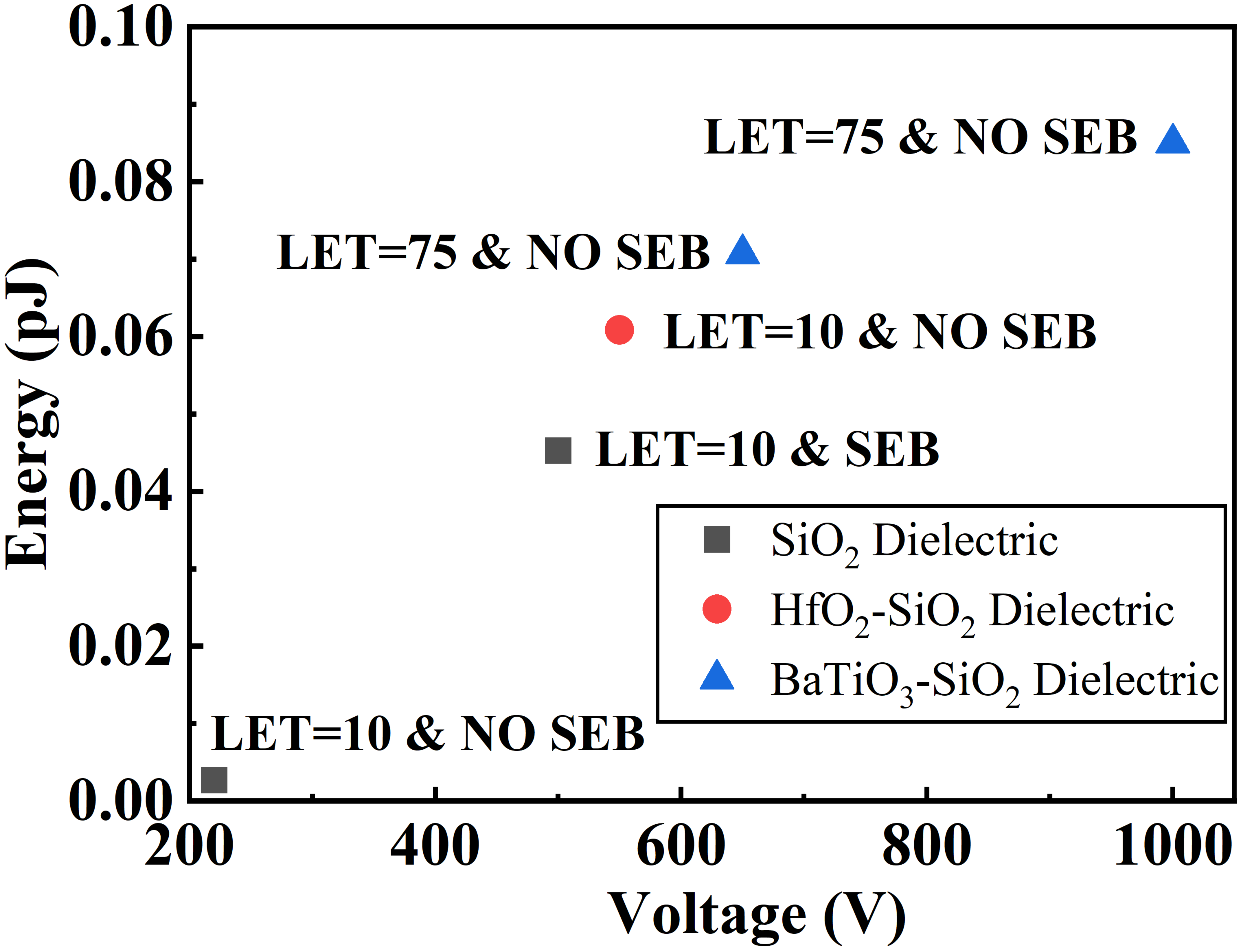}
    \caption{Energy dissipated during the ion strike event in $\beta$-Ga$_2$O$_3$ MOSFET for various radiation conditions. LET values are measured in MeV/mg/cm$^2$ }
    \label{fig15}
    \end{center}
\end{figure}
Fig.15 shows the energy dissipated for the first 336ps of the ion strike event in $\beta$-Ga$_2$O$_3$ MOSFETs under different radiation conditions. It is observed that the power dissipated for ion strike event in Ga$_2$O$_3$ MOSFET irrespective of the radiation condition is significantly less than that energy dissipated in SiC MOSFETs\cite{ball2019ion}. This results from the lower ionization rate in Ga$_2$O$_3$ due to its large bandgap. The energy dissipated increases with increasing LET and increasing operating bias which is in accordance with our discussion in the earlier sections that with increasing bias and LET, the electric fields could be higher and the transient drain current peaks are higher. However, the energy dissipation factor is not correlated with specific SEB effect as seen in the figure.
\section{Peak Electric Field and SEB Threshold condition}
\begin{figure}[h]
\begin{center}
    \includegraphics[width=\linewidth]{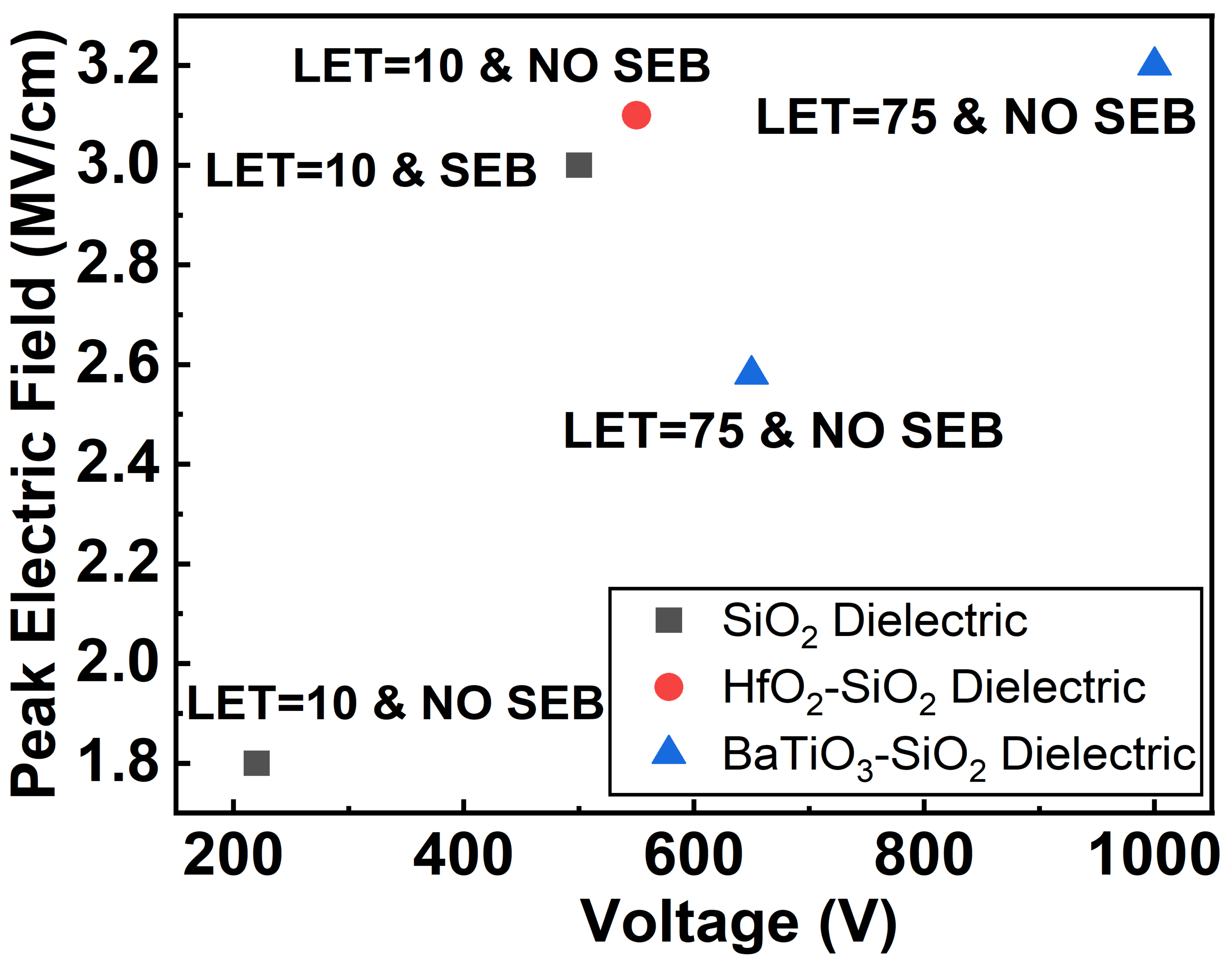}
    \caption{Peak Electric fields in the channel at t=8$\times$10$^{-9}$s in the ion strike location for all designs }
    \label{fig16}
    \end{center}
\end{figure}

As discussed in the earlier section, though the energy dissipation in Ga$_2$O$_3$ MOSFET is lower than that of SiC MOSFETs, it is not a helpful parameter in determining a specific SEB threshold condition for voltage or LET. The main factor responsible for triggering the SEB mechanism was found to be the electric field. Though our simulations suggest that decreasing the electric field in the channel increases the SEB threshold conditions, it is difficult to identify a specific threshold condition that is applicable for Ga$_2$O$_3$ MOSFETs across all design modifications. Fig.~\ref{fig16} shows the variation of peak electric field in the channel at the location of ion strike at t=8$\times$10$^{-9}$s for all the designs. It can be clearly seen that using a high-k dielectric reduces the peak electric fields in the channel but it also changes the overall electric field distribution in the device. But the peak electric field value is not the decisive factor for determining the SEB triggering mechanism. The overall electric field distribution affects the total impact generation rate in the device which is responsible for the generation of carriers. For example, if we take radiation condition of V$_{ds}$=500 V and LET=10 MeV/mg/cm$^2$ for the conventional design shown in Fig.~\ref{fig1} and a radiation condition of V$_{ds}$=1000 V and LET=75 MeV/mg/cm$^2$ for the modified design shown in Fig.~\ref{fig8} with the BaTiO$_3$ dielectric, the peak electric fields in the channel for the BaTiO$_3$-SiO$_2$ design is higher as shown in Fig.~\ref{fig16}. However, the field distributions are different for the two designs, which leads to different impact generation rates and different levels of electron and hole concentrations as discussed earlier. Thus even though the peak electric field value is higher the BaTiO$_3$-SiO$_2$ design combination does not suffer from SEB, whereas in the SiO$_2$ design, SEB is triggered.
\section{Conclusion}
In this article, detailed 2D TCAD simulations were performed to investigate the SEEs in $\beta$-Ga$_2$O$_3$ MOSFETs. The physics behind the SEB mechanism is understood and the electric field distribution in the channel is one of the main factors behind the SEB triggering mechanism. Our initial simulations suggest that the SEB threshold voltage of the baseline lateral Ga$_2$O$_3$ MOSFET is lower than the simulated SEB threshold in state of the art AlGaN/GaN HEMTs and MISFETs. Thus keeping in mind the physics behind the SEB mechanism radiation hardening techniques are proposed which show improved radiation hardened performance. To reduce the high electric fields in the channel responsible for the high impact generation rates, a radiation hardened device with rounded gates and high-k dielectric was proposed. The high-k dielectric is used in combination with SiO$_2$ to reduce the fabrication challenges associated with depositing a thick layer of high-k dielectric. Firstly, with HfO$_2$-SiO$_2$ dielectric combination, SEB threshold voltage of 550V is obtained which is higher than that of GaN HEMTs. However, this design also fails under extreme radiation conditions. Therefore it is proposed that to operate under extreme radiation conditions, a very high-k dielectric material like BaTiO$_3$ could be used which reduces the overall electric field distribution in the channel significantly. With this design, the device can achieve SEB thresholds going up to 1000V at LET=75 MeV/mg/cm$^2$. Another technique that could be explored to design radiation hardened $\beta$-Ga$_2$O$_3$ MOSFETs, is using a p-type material like p-NiO, which is also helpful in reducing the electric field in the channel.
\begin{acknowledgments}
We acknowledge the support from AFOSR under award FA9550-18-1-0479 (Program Manager: Ali Sayir), from NSF under awards ECCS 2019749, ECCS 2231026
\end{acknowledgments}
\nocite{*}
\bibliography{aipsamp}

\end{document}